\documentclass[journal]{IEEEtran}
\usepackage{booktabs}
\usepackage[justification=centering]{caption}
\usepackage{tabularx}
\usepackage[numbers,sort&compress]{natbib}

\ifCLASSINFOpdf
  \usepackage[pdftex]{graphicx}
\else
\fi
\IEEEpubid{\begin{minipage}{\textwidth}\ \centering
		Copyright \copyright 2025 IEEE. Personal use of this material is permitted. 
        Permission from IEEE must be obtained for all other uses, in any current or future media, including reprinting/republishing this material for advertising or promotional purposes, creating new collective works, for resale or redistribution to servers or lists, or reuse of any copyrighted component of this work in other works.
\end{minipage}}

\begin{document}
%
\title{LLM Enabled Multi-Agent System for 6G Networks: Framework and Method of Dual-Loop Edge-Terminal Collaboration}
%
%

\author{Zheyan Qu,
        Wenbo Wang,        
        Zitong Yu,
        Boquan Sun,
        Yang Li,
        Xing Zhang
\thanks{Zheyan Qu, Wenbo Wang (corresponding author), Zitong Yu, Boquan Sun, Yang Li, and Xing Zhang (corresponding author) are with Beijing University of Posts and Telecommunications, China.
}
}

%
%

\markboth{Journal of \LaTeX\ Class Files,~Vol.~14, No.~8, August~2015}%
{Shell \MakeLowercase{\textit{et al.}}: LLM Enabled Multi-Agent System for 6G Networks: Framework and Method of Dual-Loop Edge-Terminal Collaboration}
%



\maketitle

\begin{abstract}
The ubiquitous computing resources in 6G networks provide ideal environments for the fusion of large language models (LLMs) and intelligent services through the agent framework. With auxiliary modules and planning cores, LLM-enabled agents can autonomously plan and take actions to deal with diverse environment semantics and user intentions. However, the limited resources of individual network devices significantly hinder the efficient operation of LLM-enabled agents with complex tool calls, highlighting the urgent need for efficient multi-level device collaborations. To this end, the framework and method of the LLM-enabled multi-agent system with dual-loop terminal-edge collaborations are proposed in 6G networks. Firstly, the outer loop consists of the iterative collaborations between the global agent and multiple sub-agents deployed on edge servers and terminals, where the planning capability is enhanced through task decomposition and parallel sub-task distribution. Secondly, the inner loop utilizes sub-agents with dedicated roles to circularly reason, execute, and replan the sub-task, and the parallel tool calling generation with offloading strategies is incorporated to improve efficiency. The improved task planning capability and task execution efficiency are validated through the conducted case study in 6G-supported urban safety governance. Finally, the open challenges and future directions are thoroughly analyzed in 6G networks, accelerating the advent of the 6G era.

\end{abstract}

\begin{IEEEkeywords}
6G networks, multi-agent system, large language model, AI agents, edge computing.
\end{IEEEkeywords}

%
\IEEEpeerreviewmaketitle

\section{Introduction}
%
%
%
%

With the introduction of native intelligence, self-operation, and self-evolution are gaining increasing consensus as critical attributes of 6th generation wireless systems (6G) \cite{IEEEhowto:1}. Within the service paradigm of 6G, advancements in artificial intelligence (AI) are deeply integrated into network elements. Leveraging extensive network data, AI models can be continuously updated and collaborated to accommodate dynamically evolving network and service demands, thereby achieving a highly autonomous and intelligent network system. In this context, the traditional network services, characterized by passive predefined workflow, are progressively transforming into proactive intelligent services that spontaneously adjust service orchestration and network configuration in response to real-time contextual semantics and requests \cite{IEEEhowto:2}. 

As pivotal milestones in artificial intelligence, the Large Language Models (LLMs) are believed to serve as the bedrock for realizing the loops in self-operation and self-evolution. Owing to their exceptional perception and reasoning abilities, LLMs are poised to function as the “brain” of 6G networks. Pre-trained on extensive corpora, LLMs have exhibited remarkable performance in following user instructions and generalization across a wide range of tasks, enabling them to navigate through various complex network scenarios. In the future, the multimodal large language models (MLLMs) with communication-related modalities such as spectrum, signals, and point cloud will be further developed, better supporting 6G management. Moreover, by leveraging the reasoning capabilities of LLMs, the target intention and scene semantics can be further explored and transformed into corresponding service orchestrations, thereby enabling intelligent, on-demand services. As a significant step towards Artificial General Intelligence (AGI), the advent of LLMs offers a valuable opportunity for the evolution of 6G networks into truly intelligent entities\cite{IEEEhowto:4}.

\begin{figure}[!t]
    \centering
    \includegraphics[width=2.7in]{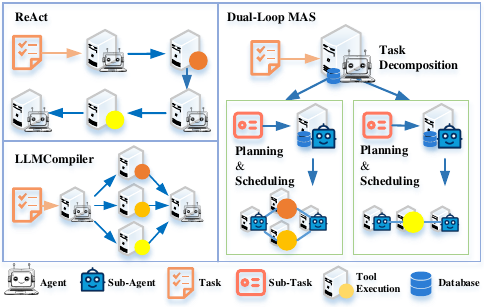}
    \caption{Overview of the dual-loop multi-agent system, compared with classical LLM-enabled agent systems.}
    \label{fig:pic0}
\end{figure}

\IEEEpubidadjcol

However, relying solely on LLMs cannot support the full operation of 6G services. Firstly, several inherent limitations of LLMs constrain their performance in operating 6G services. For instance, LLMs' deficiencies in mathematical calculations strictly impede them from network optimization tasks requiring precise numerical operations. Secondly, the substantial training resources required by LLMs severely restrict their flexibility, which makes it challenging to rapidly optimize LLMs for specific scenarios, hindering the vision of self-evolving 6G networks. Lastly, the current cloud-based LLM interfaces result in a disconnect between large models and the native intelligence in 6G networks, in which case, all the  execution details generated by intelligent services need to be uploaded to the cloud to access LLM services, resulting in significant bandwidth burdens and privacy leakage risks.

In light of these challenges, researchers are increasingly turning to the LLM-enabled Multi-Agent System (MAS). Leveraging LLMs as the planning module, MAS can intelligently analyze user requirements and execute a series of tasks through multi-agent collaborations. Each agent is equipped with various toolkits, such as calculators and network detectors, which can be called to engage with the 6G network. Moreover, external modules such as vector databases and knowledge graphs can be dynamically updated to fill the gaps in network knowledge, reducing dependence on retraining. Lastly, the distributed nature of MAS enables agents to be dispersed across different locations in the network, fostering excellent scalability.

Based on the above findings, an LLM-enabled MAS with dual-loop edge-terminal collaborations is proposed to enhance its performance in handling complex tasks and execution efficiency, facilitating the deep integration of AI with 6G networks. As illustrated in Fig. \ref {fig:pic0}, the dual-loop structure is utilized, where the outer loop consists of task decomposition and the inner loop consists of LLMCompiler \cite{IEEEhowto:5}. The requests from 6G network users are automatically analyzed and decomposed into subtasks by the planning of the global agent in the outer loop. Subsequently, these parallel subtasks will be handed over to a series of sub-agents with designated roles to generate the parallel execution details with LLMCompiler. Moreover, the network functions are further integrated as part of the toolkits to enable the fusion of MAS and scheduling in 6G networks, where agents can determine the calling dependency of tools and orchestrate computing resources through offloading functions. Finally, with the memory module, the proposed system can leverage stored experiences and facilitate self-evolution. As the sub-task planning and tool execution are all in parallel, the overall efficiency can be effectively improved, and the system can be easily applied to different scales, enhancing overall flexibility.

With the ubiquitous connectivity and multi-level computing resources of 6G networks, MAS can orchestrate various network services and toolkits through collaborations to meet dynamic user demands. Briefly, the principal contributions of this article are summarized as follows:

\begin{itemize}
  \item The potential of LLM-enabled agents to realize 6G functions is summarized and discussed, uncovering existing research gaps.
  \item An LLM-enabled MAS with dual-loop edge-terminal collaboration is proposed for 6G network services, with a case study in 6G urban safety governance to validate the improved planning capability and task execution efficiency.
  \item The open challenges and future directions are analyzed, paving the way for deeply integrating LLM-enabled agents in the 6G era.


\end{itemize}

\begin{table*}[!htbp]
\centering
\caption{LLM-enabled agent functions in 6G features.}
\begin{tabularx}{\textwidth}{|l|X|l|X|}
\toprule
Agent Functions & 6G Features & Ref. & Research Gaps \\
\midrule
Perception and Cognition & Semantic Communication, Intent-driven Network& \cite{IEEEhowto:6, IEEEhowto:7} & Hallucination in Multi-modal Fusion and Intention Recognition
\\
\hline 
Orchestration & Integration of Full Resources, Cloud/fog/edge Computing & \cite{IEEEhowto:8} & Constrained Reasoning Ability and Network Analysis Capabilities \\ 
\hline 
Self-adaption and Self-evolution & Digital Twins, Self-evolving Network & \cite{IEEEhowto:10, IEEEhowto:11} & High Natural-language-based Self-optimization Complexity\\
\hline 
Knowledge Management & Automated Network Management & \cite{IEEEhowto:14} & Need for 6G-oriented Pretraining \\
\bottomrule
\end{tabularx}%
\label{tab:6G}%
\end{table*}%

\section{LLM-enabled Agent system for 6G Networks}
The LLM-enabled Agent system, as an integration of AI and various classical applications, aligns seamlessly with the concept of AI-supported 6G networks. Specifically, it plays a significant role in the following aspects of 6G networks, which are summarized in Table \ref{tab:6G}.


\subsubsection{Perception and Cognition} The future 6G networks are expected to provide customized services that meet highly dynamic scenarios, requiring the perception of multimodal information with semantic and intention information.

Equipped with sensors of different modalities, the LLM-enabled agent is capable of perceiving, aligning, and analyzing the feature spaces of different modalities. For instance, in multimodal semantic communication, pre-trained LLMs, visual encoders, and deep fusion modules can be integrated with the agent and dynamically called to process single-modal or multi-modal inputs\cite{IEEEhowto:6}, which will then be transmitted to the planning module for further processing.

Moreover, the perception of user intention plays a crucial role in 6G services, which necessitates a deeper step on perception results. Based on the robust natural language understanding and reasoning abilities of LLMs, the agent can deconstruct semantic information and generate intention parameters. In agent-based 6G intent-driven networks, users only need to suggest desired states, and the service intents will be analysed by multiple agents' negotiations to generate network intents for network slicing\cite{IEEEhowto:7}.

Though agents can achieve high-level perception and cognition functions, they still face great challenges in hallucinations, which may be caused by insufficient fusion blocks or disorganization between multiple agents. In this case, the coordination of MAS with perception modules still needs further exploration.



\subsubsection{Orchestration} Facing global coverage visions in 6G networks, an intelligent resource scheduling system towards multi-tasking and large-scale connectivity is of great significance. First, LLM-enabled agents can decompose complex requests into multiple subtasks and delegate them to different devices. Different from traditional workflows, where network services are predefined and fixed, agent-based service orchestration is dynamic and on-demand. By deploying these agents, along with toolkits, on mobile devices and edge servers, the global collaboration of multi-level resources will be further promoted \cite {IEEEhowto:8}. 

Furthermore, network optimization models can be integrated as part of the toolkit, including various metaheuristic algorithms. As these algorithms usually have different properties in complexity, energy efficiency, latency efficiency, etc., the agent has to first analyze user requirements, network environment, and algorithm properties, before adaptively calling proper network optimization tools\cite{IEEEhowto:9}. In response to diverse scenarios, such as latency-sensitive tasks or energy-sensitive tasks, the agent should intelligently match the algorithm properties with specific contexts, thereby enabling the automation of network operations and optimization.

However, managing multi-level resources also brings about higher requirements of task planning and network analysis capabilities, to solve which the architecture of multi-agent reasoning necessitated further research.

\subsubsection{Self-adaption and Self-evolution} The rapid evolution of user demands and networks poses a significant challenge to the adaptability of future communication systems, in which case network services need to be constantly updated and evolved. LLMs acquire world knowledge and robust generalization capabilities after training with massive amounts of data, which enables LLM-enabled agents to support flexible adaptation of 6G networks. Based on digital twins, the LLM-enabled agent can automatically interact with the simulation and explore feasible parameters through natural language reasoning, thereby accommodating the changing environment \cite {IEEEhowto:10}.

Furthermore, this versatility endows LLM-enabled agents with the capability to self-evolve without human intervention. The successful experience can be automatically summarized and stored as memory, which can be promptly retrieved when handling relevant contexts. For instance, to improve the user experience, the agents can first optimize their profiles based on the user requirements instead of adopting the predefined structure, and then constantly update their roles according to feedback\cite {IEEEhowto:11}. Concurrently, agents can further fine-tune themselves through methods like Low-Rank Adaptation (LoRA) to improve their reasoning ability in certain scenarios, where the updated small amounts of parameters can be stored separately and dynamically loaded \cite{IEEEhowto:12}. 

Overall, through the above methods, LLM-enabled agents are promised to achieve self-adapted and self-evolved 6G services. Correspondingly, as natural-language-based self-optimization generally involves multi-turn dialogue, the mechanism should be elaborately designed to reduce complexity.

\subsubsection{Knowledge Management} LLMs play a pivotal role in managing and utilizing knowledge in 6G networks. With the wide use of LLMs, 6G networks are poised to support a series of knowledge-oriented conversational services. 

LLM-enabled agents not only encompass the general world knowledge but also can integrate various knowledge retrieval modules, such as databases and search tools. By extracting knowledge from professional documents, literature data, and history planning paths, to guide future actions \cite {IEEEhowto:14}, the agent can further expand its ability to manage network status.

Vice versa, LLM-enabled agents also serve as a bridge that facilitates the conversion of 6G knowledge into applications. Based on the understanding of knowledge, these agents can not only generate various strategies but also translate them into physical tool calls\cite {IEEEhowto:14}. For instance, in automated network operation and optimization, agents can acquire network status and retrieve relevant operation knowledge from the database for analysis, which can be subsequently used to call a range of network tools to recover and improve the network setting. 


In MAS, the knowledge management module should also be integrated to improve each agent's abilities. Furthermore, aside from plug-in modules, the 6G-oriented LLMs should also be developed with pretraining on 6G-related data, where the knowledge will be inherent in LLMs' parameters.

\begin{figure*}[!t]
    \centering
    \includegraphics[width=4.6in]{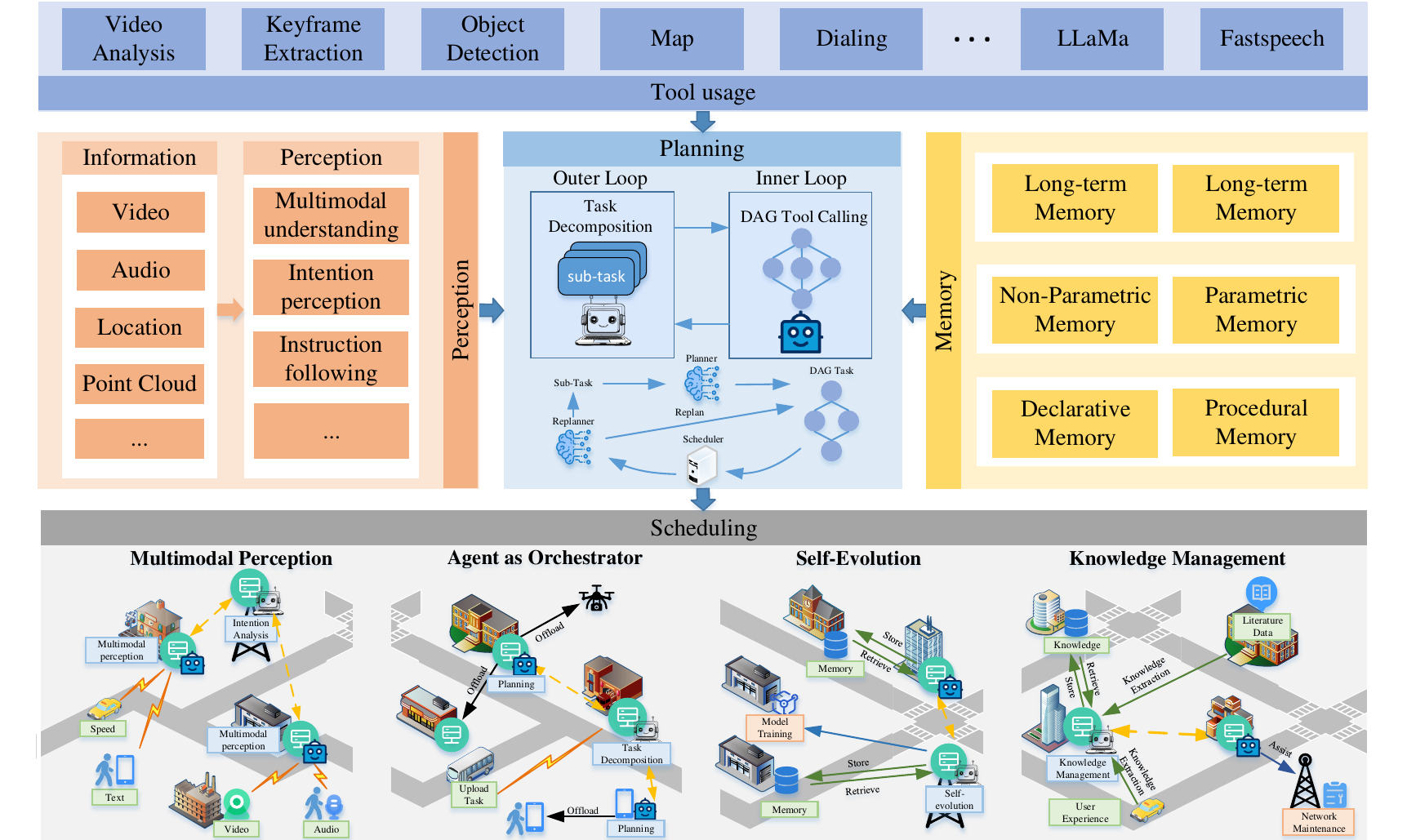} 
    \caption{The dual-loop multi-agent system with terminal-edge collaboration for efficient 6G AI applications.}
    \label{fig:pic1}
\end{figure*}

\section{Dual-loop Multi-Agent Framework with Parallel Terminal-Edge Collaboration}

Targeted at 6G visions, we developed a dual-loop multi-agent framework with terminal-edge collaboration, where the parallel tool calls are hierarchically generated through inter-agent collaboration and intra-agent planning. Different from classical agent schemes, the proposed framework further incorporates service scheduling to offload tool executions to multi-level devices. The overall structure is illustrated in Fig. \ref {fig:pic1}.

\subsection{Multimodal Perception}

To provide dynamic, on-demand intelligent services for 6G users, LLM-enabled agents must first perceive multimodal environmental information and user intentions. Specifically, the perception module is responsible for the following functions:

\subsubsection{Multimodal information perception} The complementarity between multimodal information can enhance the perception accuracy. In 6G services, there may be signals of diverse modalities. Given the potential noise in the data, it is essential to first denoise and align these multi-modal data, converting them into a standard format manageable by the agent. This process can be accomplished through AI models or specialized classic algorithms, such as video denoising models and signal filtering. Moreover, the perception module is also a crucial interface for interaction between the user and the LLM-enabled agent, where users can provide natural language instructions with intentions.

\subsubsection{Intention recognition} As an integral component of 6G intelligent services, extracting scene semantics and user intentions constitutes a critical step. On the one hand, in scenarios without explicit user instructions, the perception module is required to autonomously identify scene semantics according to configurations and guide the planning module in executing tasks. For instance, in automated urban emergency responses, upon detecting anomalies, the perception module must promptly formulate appropriate instructions, such as carrying out first aid and setting traffic warnings. 

On the other hand, when user instructions are appointed, the perception module serves as the interactive interface. In LLM-enabled 6G network management, diverse network infrastructures are formulated as declarative intents, where user instructions must be analysed and converted into network parameters \cite{IEEEhowto:7}.

\subsection{Memory}

The memory module is a crucial component responsible for storing the interaction context, 6G knowledge, and action experiences to guide its decision-making processes. By summarizing operation trajectories and extracting network knowledge, the agent can continuously refine and enhance its reasoning capabilities to achieve self-evolution. Though the potential overlaps, the memory module can generally be divided into the following sections:

\subsubsection{Short-term memory / Long-term memory} Short-term memory stores real-time context and input data required for decision-making, such as the reasoning history. Long-term memory records 6G network knowledge and action experiences. For instance, in network operation and maintenance, the agent can retrieve the network operations manual to deal with a series of network anomalies, and the planning trajectories can be summarized as action experiences for future reference.

\subsubsection{Non-parametric memory / Parametric memory} Non-parametric memory is primarily stored in original formats, such as textual beliefs, images, and signals, which can be embedded into vectors and easily retrieved. Parametric memory mainly refers to the parameter matrices retained by self-fine-tuning, such as the low-rank matrix generated from LoRA \cite{IEEEhowto:12}, where this portion of parameters can be rapidly loaded without the necessity of reloading the entire model.

\subsubsection{Declarative memory / Procedural memory} The concepts of declarative memory and procedural memory are inspired by research on the human memory system. Declarative memory primarily stores factual knowledge and concepts, such as user preferences and profiles. In contrast, procedural memory is more oriented toward the storage of "skills," such as strategies for agents to take action in specific scenarios. By integrating these two types of memory, an agent can closely mimic the human memory system, thereby delivering high-quality services akin to those provided by human experts.

\begin{figure*}[!htb]
    \centering
    \includegraphics[width=4.8in]{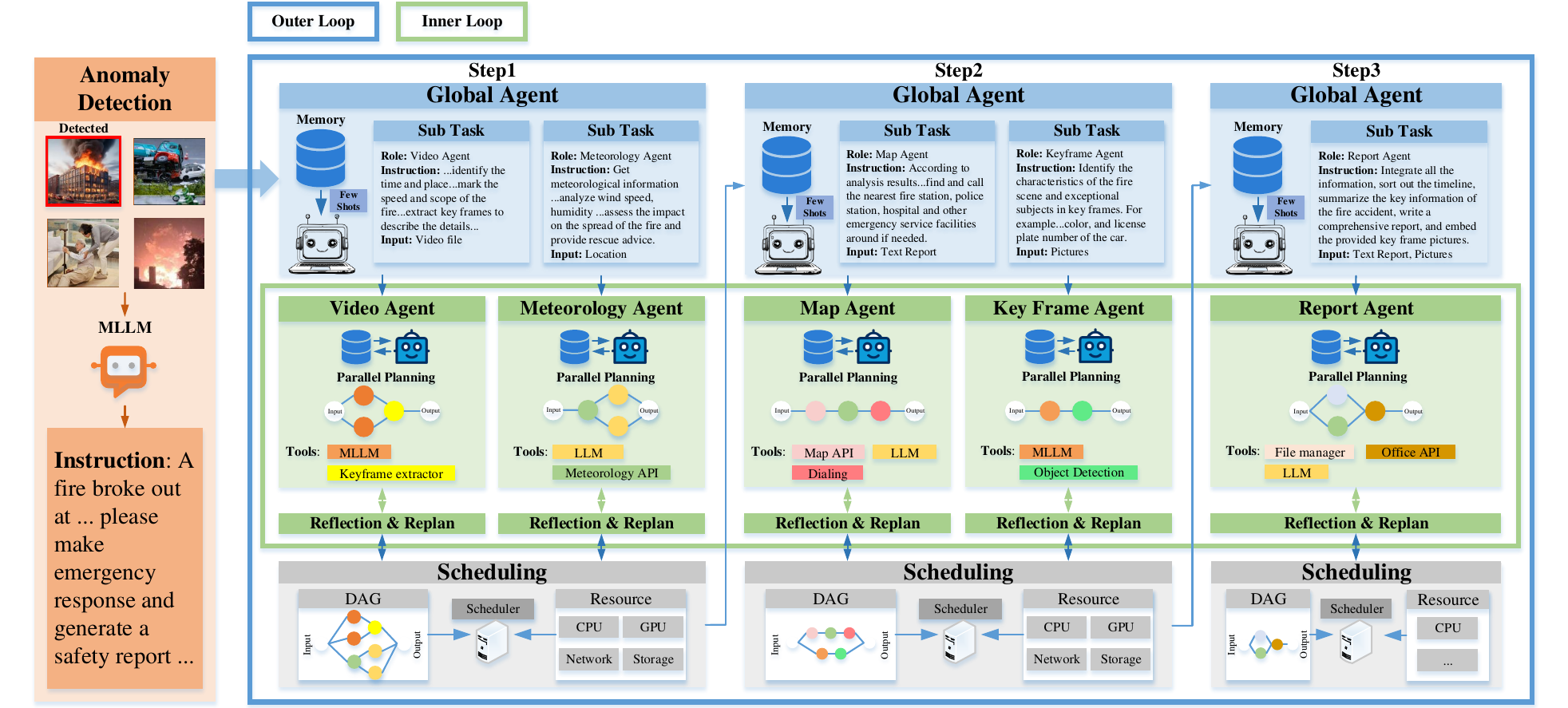} 
    \caption{One case of parallel task planning in the dual-loop multi-agent framework within urban emergency response tasks.}
    \label{fig:pic2}
\end{figure*}

\subsection{Planning} 

Serving as the core component of the agent, the planning module is responsible for adhering to instructions and planning a series of tasks based on perceived information.  To improve reasoning performance and reduce hallucination, a dual-loop structure is introduced with the LLMCompiler, where the parallelism in both the multi-agent cooperation and tool calling is further enhanced with edge-terminal collaboration. The specific design is as follows:

\subsubsection{Outer loop} Confronted with complex tasks, traditional agent systems often encounter limitations in reasoning ability and context window size. In this case, the model is prone to losing critical contextual information, leading to planning hallucinations. To address this issue, the designed outer loop first decomposes the task into multiple parallel subtasks by the global agent on edge devices, and subsequently assigns these subtasks to sub-agents with different roles on terminals. Once the sub-agents complete their assigned subtasks, the outcomes are aggregated to the global agent for the next-step subtasks generation. 

Through task decomposition, the original complex task will be converted into multiple parallel subtasks, which can be handed over to multiple sub-agents to improve execution efficiency. Moreover, as subtasks generally involve fewer reasoning steps and lower reasoning capability, relatively smaller parameter amounts are needed for sub-agents. Lastly, the hierarchical framework allows for scalable and independent deployment across different devices and regions, where each edge-side global agent covers the task decomposition and planning in a certain area, and each sub-agent with independent subnets accesses local data to complete subtasks. Therefore, the safe isolation and excellent scalability can be achieved.

\subsubsection{Inner loop} In the inner loop, multi-role sub-agents are employed to complete subtasks by calling functions and tools, fostering edge-terminal device collaborations. Independent terminal-side sub-agents can be freely placed at on-demand locations to access nearby sensors and toolsets. After sub-task planning, the network operation functions, such as offloading models, will be called to orchestrate toolset execution. Meanwhile, leveraging the role-playing capabilities of LLMs, these agents can improve their performance within specific contexts. For instance, an agent with the role of network manager can simulate the decision-making process of an experienced professional, applying best practices to improve network performance. 

Moreover, the LLMCompiler is integrated to generate the parallel tool calling to reduce task execution latency. In most current implementations, the generated tool calling in 6G networks remains sequential, where only one tool will be executed in each iteration \cite{IEEEhowto:15}. These schemes overlooked the parallelism inherent in subtasks. In the proposed system, sub-agents of different roles first generate a parallel tool calling topologies according to the sub-task instruction, where the dependency between multiple tools consists of a Directed Acyclic Graph (DAG), and then dispatch them to appropriate computing nodes for execution. Subsequently, the sub-agents dynamically adjust the DAG structure based on the execution results and feedback from the tools, which is known as replanning. This approach not only increases the overall parallelism of task execution but also fosters collaboration among devices across the network.

\subsection{Scheduling} 

In 6G networks, agent applications should provide on-demand and personalized services while achieving a profound integration with the network system. Following the planning of parallel tool calling, the execution of these programs should be further offloaded to ubiquitous computational resources.

Agents can encompass various kinds of tools, including classic algorithms, machine learning models, or even LLMs, which require different levels of computing resources. Concurrently, terminals and edge devices in 6G networks are manufactured for various purposes and thus possess heterogeneous computing resources. Therefore, it is essential to carefully match the intricate relationship between the resource demands of tools and the supply of devices to facilitate reasonable scheduling, enhancing task execution efficiency.

However, constrained by inherent limitations, it is challenging to achieve highly efficient scheduling and precise resource division only by applying LLMs. In this context, the scheduling algorithms are incorporated as part of the toolkits and enable agents to intelligently select suitable algorithms based on different business and network states. For instance, applications involving different transport protocols require different transmission modeling, where agents can adaptively choose the appropriate algorithms in real time according to the application's settings.

\section{Case Study}

To validate the proposed framework in 6G networks, a case study is conducted within the urban emergency response scenario focused on the task planning capability and task execution efficiency. In this simulation, agents can detect anomalous behaviors in the city through image sensors and execute a series of automated emergency management tasks via multi-agent collaboration. 

As shown in Fig. \ref {fig:pic2}, leveraging the multimodal perception of MLLMs, the video signals of anomaly events are detected, and corresponding instructions are generated. In each iteration, the global agent first receives the information to perform task decomposition, where the generated subtasks are allocated to sub-agents. These sub-agents then leverage related experiences as few shots, and generate the parallel tool calling with DAG topologies. Lastly, the scheduling algorithms are utilized to offload tool executions to edge and terminal devices. During execution, the sub-agent continuously collects feedback for reflection and replanning to avoid planning hallucination. 

In our experiments, distinct roles are assigned to sub-agents for different subtasks, including Video-Agent, Meteorology-Agent, Map-Agent, Keyframe-Agent, and Report-Agent. As illustrated in Fig. \ref {fig:pic2}, a set of toolkits is also integrated, such as object detection, map, and file managers. Moreover, to comprehensively test the performance of the proposed system, an emergency response task set is designed with varying difficulty levels. The task set involves video clips of multiple anomaly scenarios, such as fire and traffic accident, along with corresponding instructions of different complexities, such as capturing keyframes, initiating distress calls, and generating accident reports. The difficulty level of each task is determined by the number of tool calls needed and is categorized into easy (1-3 tool calls), medium (4-6 tool calls), and hard (7-9 tool calls). In practice, the MiniCPM-V 2.6 \cite{IEEEhowto:16} hosted on A40-48GB GPU is employed for video understanding, and the model for agent planning is developed based on GLM-4-0520 \cite{IEEEhowto:17} with 4095 max tokens and zero temperature.

\begin{figure}[!t]
    \centering
    \includegraphics[width=2.7in]{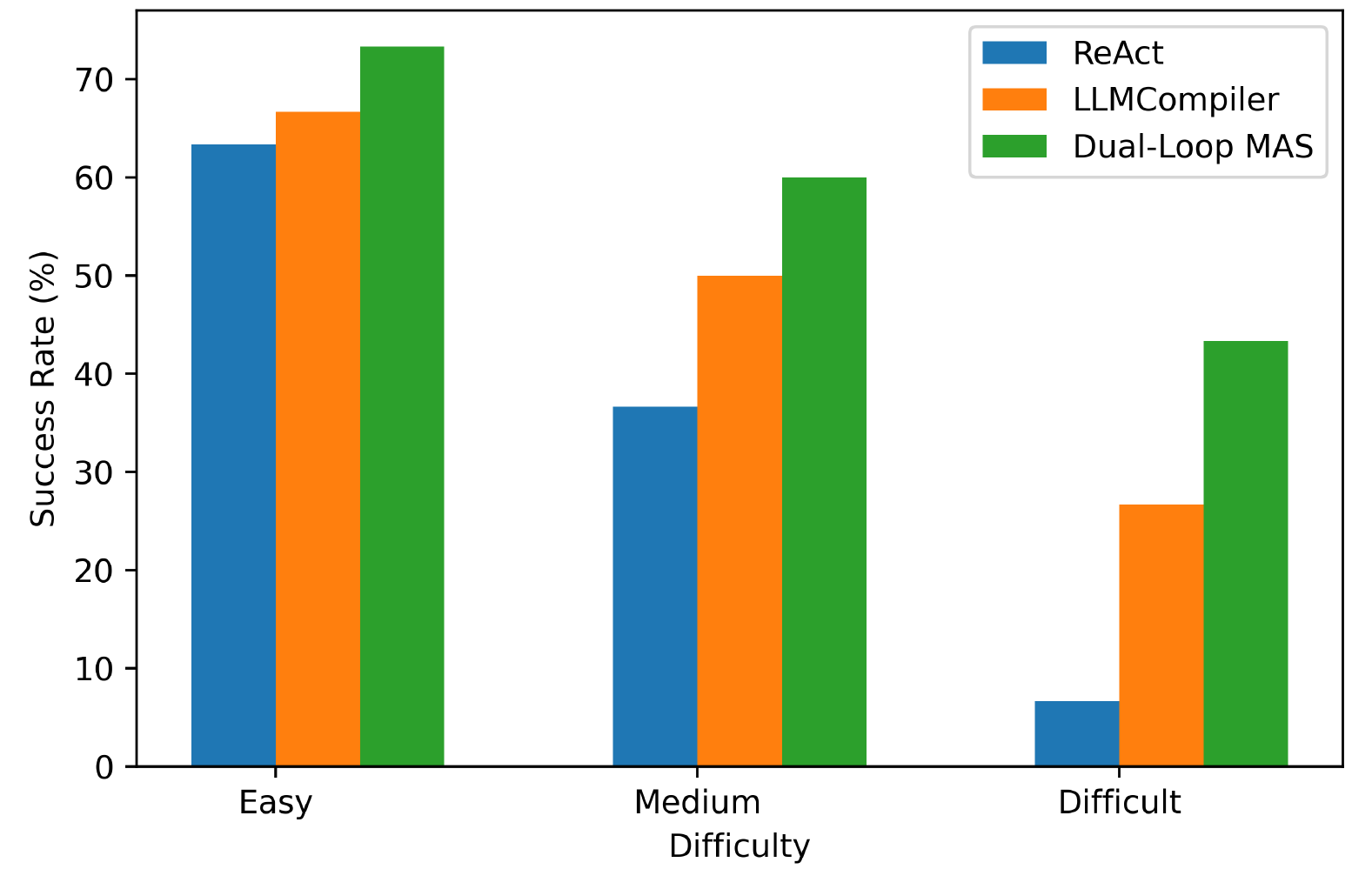} 
    \caption{Success rate on tasks with different difficulties.}
    \label{fig:pic3}
\end{figure}


\begin{figure}[!t]
    \centering
    \includegraphics[width=2.85in]{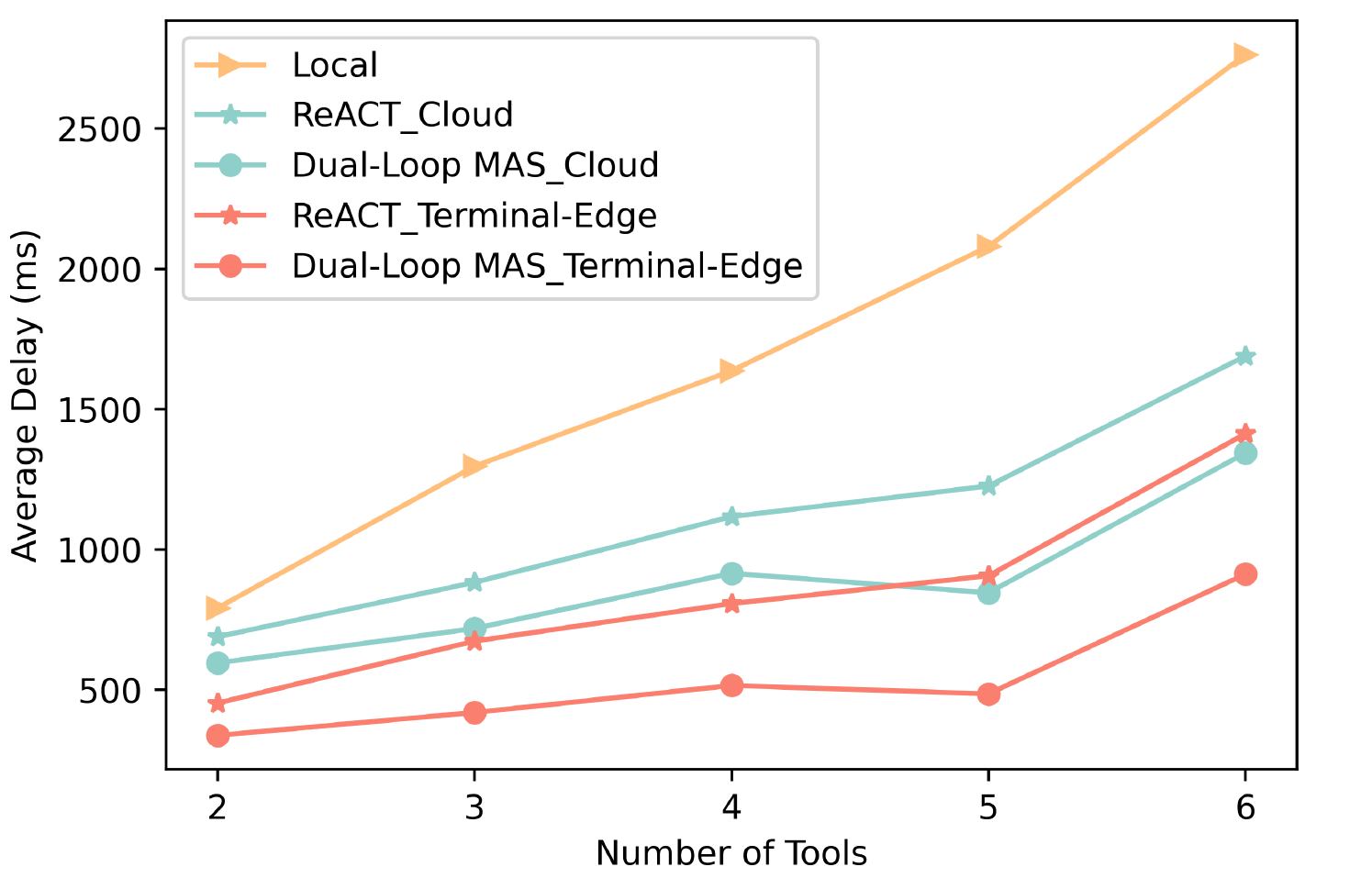} 
    \caption{Latency by the different number of tools called.}
    \label{fig:pic4}
\end{figure}

The experimental results of success rate (SR) are illustrated in Fig. \ref {fig:pic3}, which evaluates the percentage of successfully planned tasks. Classical agent schemes, including ReAct and LLMCompiler\cite{IEEEhowto:5, IEEEhowto:15}, are set as baselines, whose planning processes are shown in Fig. \ref {fig:pic0}. As task complexity increases, the SR of all three schemes exhibits varying degrees of decline, which is attributed to the increased reasoning difficulty and context length. Especially, as the planning process of ReAct is sequential, it quickly encounters early stops, where the planning process may prematurely terminate before the completion \cite{IEEEhowto:15}. In contrast, the proposed system achieves the best performance. This is because it first decomposes the task into multiple simpler subtasks, which significantly reduces the difficulty of sub-task planning for each sub-agent. Meanwhile, leveraging related experiences and demonstrations, the reasoning hallucinations are also mitigated.

Furthermore, the execution latency of the tool execution across different schemes is compared in Fig. \ref {fig:pic4}, where the latency is calculated based on the generated tool topology. In our simulation, 5 edge servers and 10 terminals with heterogeneous computational resources are set for collaboration. Note that only tasks with successful planning and execution are considered, and the priority-based algorithm is chosen as the scheduling tool, where each tool is sorted by the length of the critical path and scheduled in order. Compared with the terminal-edge collaboration scheme with cloud execution and local execution, it is found that utilizing multi-device collaboration can effectively reduce latency. Moreover, as the tool execution topology of the proposed system is parallel and that of ReAct is sequential \cite{IEEEhowto:15}, the latency is further reduced both in cloud execution and terminal-edge collaboration.

Lastly, it is noteworthy that the proposed framework can be freely extended to other 6G scenarios. For instance, in network slicing management, sub-agents with corresponding roles will be responsible for each key process, such as intent analysis, resource coordination, and lifecycle management, and calling various slice modeling and deployment toolsets to complete subtasks. Meanwhile, the global agent will generate overall task planning and decomposition based on user requirements to orchestrate these sub-agents. Similarly, this paradigm can be adapted to scenarios like satellite-ground communication, digital twins, etc.

\section{Open Challenges and Future Directions}

Although deploying LLM-enabled agents on multi-level networks is a promising way to achieve 6G intelligent services and native AI, there are still several open challenges.

\subsubsection{Flexible On-Device Deployment} In most current applications, the core LLMs are deployed on the cloud, where all the interaction details have to be uploaded, and user privacy cannot be guaranteed. Though the cloud is irreplaceable as it maintains the most powerful closed-source LLMs, the flexible on-device deployment should not be overlooked, which is critical for native intelligence and local private services. 

However, limited resources of terminals and edge servers significantly restrict the performance of LLM-enabled agents, as more advanced general abilities generally involve larger parameter amounts. Therefore, it's significant to develop on-device LLMs tailored for 6G network operations and task planning, where more 6G network knowledge should be incorporated in pretraining. Besides, based on on-device models, the balance between LLM capability requirements and resource consumption should be achieved to fully leverage the collaboration of multi-level devices in MAS.

\subsubsection{Reasoning Ability and Context Window} LLM-enabled agents can surmount the inherent limitations by calling external tools, which also pose great challenges to reasoning capabilities and the context window size. Generally, agents are required to engage in dozens of steps of reasoning to determine the necessary tool calling and corresponding parameters. Nevertheless, the limited context window and intrinsic reasoning deficiencies strictly restrict its effective task planning. Though MAS can improve task planning capabilities to some extent, they still encounter great difficulties in dealing with complex tasks. Especially when handling long context, the base LLM often finds it difficult to extract desired details and resulting in early stop or even task failure. Therefore, there remain great challenges in improving agents' reasoning capabilities.

\subsubsection{Communication Overhead and Security} The reasoning and planning of agent systems generally involve multi-round exchanges of private data and execution parameters, which may put excessive communication overhead and arouse information leakage risk during transmission. In this case, the key information extraction strategy and the routing mechanism should be developed to reduce communication volumes and improve communication balance. Moreover, besides the proposed distributed multi-agent framework, where each agent is isolated, the encrypted communication and distributed training strategies should be incorporated to prevent privacy leakage. For instance, when developing parametric memory, the federated fine-tuning can be involved to avoid local data uploading. Overall, it's significant to improve communication mechanisms and ensure endogenous security in 6G networks.



\subsubsection{Hallucination} Hallucination indicates that the generated content may conflict with user inputs, context, or world knowledge. In 6G networks, the hallucination may lead to severe declines in user experience and network performance. Even with accurate 6G knowledge as references, the agent may still generate factually erroneous outputs. Meanwhile, the agent system may be insufficient to discover hallucinations and reasoning hallucinations promptly, which necessitates external supervisors and evaluators.  To construct reliable 6G agent applications, the hallucination mitigation method is a significant topic requiring further research.

\section{Conclusion}
In this paper, we have proposed a dual-loop edge-terminal collaboration framework for LLM-enabled MAS in 6G networks. The proposed framework leverages a dual-loop structure to improve multi-agent collaboration and reasoning capabilities. Meanwhile, the parallel tool calling scheme LLMCompiler with network scheduling is integrated, which effectively improves the multi-device collaboration and task efficiency. Moreover, we have thoroughly analyzed the ways for LLM-enabled agents to achieve 6G functions and future directions. We hope this work can provide valuable insights for future 6G developments.


\section{Acknowledgment}
This work was supported by the National Science Foundation of China under Grant 62271062. 

\ifCLASSOPTIONcaptionsoff
  \newpage
\fi

\section{Biography Section}
 

\vspace{-33pt}
\begin{IEEEbiographynophoto}{ZHEYAN QU}
(zheyanqu@bupt.edu.cn) is currently pursuing the M.S. degree at Beijing University of Posts and Telecommunications, China. His research interests include edge intelligence, multi-agent systems, and generative AI.
\end{IEEEbiographynophoto}

\vspace{-33pt}
\begin{IEEEbiographynophoto}{WENBO WANG}
(wbwang@bupt.edu.cn) is a Professor with the School of Information and Communications Engineering, Beijing University of Posts and Telecommunications, China. His current research interests include radio transmission technology, wireless network theory, cognitive communications, and software radio technology.
\end{IEEEbiographynophoto}

\vspace{-33pt}
\begin{IEEEbiographynophoto}{ZITONG YU}
(yztong@bupt.edu.cn) is currently pursuing the M.S. degree at Beijing University of Posts and Telecommunications, China. His research interests include UAV swarm coordination and multi-agent systems.
\end{IEEEbiographynophoto}

\vspace{-33pt}
\begin{IEEEbiographynophoto}{BOQUAN SUN}
(boquan\_sun@bupt.edu.cn) is currently pursuing the M.S. degree at Beijing University of Posts and Telecommunications, China. His research interests include edge computing and large language models.
\end{IEEEbiographynophoto}

\vspace{-33pt}
\begin{IEEEbiographynophoto}{YANG LI}
(ly209991@bupt.edu.cn) is currently pursuing the Ph.D. degree at Beijing University of Posts and Telecommunications, China. His research interests include device-assisted mobile edge networks, computing offloading and resource allocation.
\end{IEEEbiographynophoto}

\vspace{-33pt}
\begin{IEEEbiographynophoto}{XING ZHANG}
(zhangx@ieee.org ) is a full professor with the School of Information and Communications Engineering, Beijing University of Posts and Telecommunications, China. His research interests are mainly in 5G/6G networks, satellite communications, edge intelligence, and Internet of Things.
\end{IEEEbiographynophoto}

\vfill

\begin{thebibliography}{1}


\bibitem{IEEEhowto:1}
A. Farajzadeh \emph{et al.}, "Self-Evolving Integrated Vertical Heterogeneous Networks," \emph{IEEE Open Journal of the Communications Society}, vol. 4, 2023, pp. 552-580.
\bibitem{IEEEhowto:2}
C. -X. Wang \emph{et al.}, "On the Road to 6G: Visions, Requirements, Key Technologies, and Testbeds,"\emph{ IEEE Communications Surveys \& Tutorials}, vol. 25, no. 2, 2023, pp. 905-974.
\bibitem{IEEEhowto:4}
C. Chaccour \emph{et al.}, "Telecom’s Artificial General Intelligence (AGI) Vision: Beyond the GenAI Frontier," \emph{IEEE Network}, vol. 38, no. 5, 2024, pp. 21-28.
\bibitem{IEEEhowto:5}
S. Kim \emph{et al.},  "An LLM compiler for parallel function calling," arXiv preprint arXiv: 2312.04511, Jun 2024.
\bibitem{IEEEhowto:6}
Y. Zhao \emph{et al.}, "LaMoSC: Large Language Model-Driven Semantic Communication System for Visual Transmission," \emph{IEEE Transactions on Cognitive Communications and Networking}, vol. 10, no. 6, 2024, pp. 2005-2018.
\bibitem{IEEEhowto:7}
I. Chatzistefanidis \emph{et al.}, "Maestro: LLM-Driven Collaborative Automation of Intent-Based 6G Networks," \emph{IEEE Networking Letters}, 2024.

\bibitem{IEEEhowto:8}
M. Xu \emph{et al.}, "When Large Language Model Agents Meet 6G Networks: Perception, Grounding, and Alignment," \emph{IEEE Wireless Communications}, vol. 31, no. 6, 2024, pp. 63-71.
\bibitem{IEEEhowto:9}
A. M. Rahmani, \emph{et al.}, "Optimizing task offloading with metaheuristic algorithms across cloud, fog, and edge computing networks: A comprehensive survey and state-of-the-art schemes," \emph{Sustain. Comput. Informatics Syst.}, vol. 45, 2025, pp. 101080.

\bibitem{IEEEhowto:10}
Y. Xia \emph{et al.}, "LLM experiments with simulation: Large Language Model Multi-Agent System for Process Simulation Parametrization in Digital Twins," arXiv preprint arXiv: 2405.18092, Jul 2024.
\bibitem{IEEEhowto:11}
S. Lu \emph{et al.}, "MorphAgent: Empowering Agents through Self-Evolving Profiles and Decentralized Collaboration," arXiv preprint arXiv: 2410.15048, Oct 2024.
\bibitem{IEEEhowto:12}
E. J. Hu \emph{et al.}, "LoRA: Low-Rank Adaptation of Large Language Models," arXiv preprint arXiv: 2106.09685, Oct 2021.

\bibitem{IEEEhowto:14}
Y. Zhu \emph{et al.}, "KnowAgent: Knowledge-Augmented Planning for LLM-Based Agents," arXiv preprint arXiv: 2403.03101, Mar 2024.

\bibitem{IEEEhowto:15}
S. Yao \emph{et al.}, "ReAct: Synergizing Reasoning and Acting in Language Models," arXiv preprint arXiv: 2210.03629, Mar 2023.
\bibitem{IEEEhowto:16}
Y. Yao \emph{et al.}, "MiniCPM-V: A GPT-4V Level MLLM on Your Phone," arXiv preprint arXiv: 2408.01800, Aug 2024.

\bibitem{IEEEhowto:17}
A. Zeng \emph{et al.}, "ChatGLM: A Family of Large Language Models from GLM-130B to GLM-4 All Tools," arXiv preprint arXiv: 2406.12793, Jul 2024.
\end{thebibliography}
\end{document}